\documentclass[12pt]{article}

\usepackage[margin=1in]{geometry}
\usepackage{setspace}
\usepackage{graphicx}
\usepackage{algorithm}
\usepackage{algorithmicx}
\usepackage{algpseudocode}

\floatname{algorithm}{Algorithm}
\usepackage{url, amssymb, amsfonts}
\usepackage{amsmath}
\usepackage{amsthm}
\theoremstyle{plain}

\newtheorem{theorem}{Theorem}

\usepackage{thmtools, thm-restate}
\usepackage{xcolor}
\usepackage{multirow}
\usepackage[round]{natbib}
\usepackage{enumerate}

\begin{document}

\title{Linear Discriminant Analysis with Gradient Optimization}

\author{
Cencheng Shen\thanks{Corresponding author. Email: c.shen@microsoft.com. Microsoft Research, Redmond, WA 98052, USA.}
\and
Yuexiao Dong\thanks{Department of Statistics, Operations, and Data Science, Temple University, Philadelphia, PA 19122, USA.}
}

\date{}
\maketitle

\begin{abstract}
Linear discriminant analysis (LDA) is a fundamental classification and dimension reduction method that achieves Bayes optimality under Gaussian mixture, but often struggles in high-dimensional settings where the covariance matrix cannot be reliably estimated. We propose LDA with gradient optimization (LDA-GO), which learns a low-rank precision matrix via scalable gradient-based optimization. The method automatically selects between a Gaussian likelihood and a cross-entropy loss using data-driven structural diagnostics, adapting to the signal structure without user tuning. The gradient computation avoids any quadratic-sized intermediate matrix, keeping the per-iteration cost linear in the number of dimensions. Theoretically, we prove several properties of the method, including the convexity of the objective functions, Bayes-optimality of the method, and a finite-sample bound of the excess error. Numerically, we conducted a variety of simulations and real data experiments to show that LDA-GO wins a majority of settings among other LDA variants, particularly in sparse-signal high-dimensional regimes.
\end{abstract}


\section{Introduction}

Linear discriminant analysis (LDA) is a classification method first introduced by \citet{fisher1936use}. It assumes that data from each class are drawn from a multivariate Gaussian distribution with class-specific means and a shared covariance matrix. Under these Gaussian assumptions, LDA yields a decision boundary that is Bayes optimal \citep{hastie2009elements}. In practice, LDA estimates all parameters of the Gaussian distributions, including the class priors, class-conditional means, and shared covariance matrix, using maximum likelihood estimators. LDA has found broad application in pattern recognition \citep{mika1999fisher,duda2001pattern,blankertz2011single, dorfer2015deep, GEEDistanceGraph} and dimension reduction \citep{belhumeur1997eigenfaces,martinez2001pca,sugiyama2007dimensionality}.

The primary bottleneck of classical LDA is covariance estimation. When the number of dimensions $p$ is large relative to the number of observations $n$, the sample covariance matrix becomes ill-conditioned or singular and may deviate significantly from the true covariance matrix, such that the resulting classifier can perform poorly even when the Gaussian assumptions hold. To address this, several approaches have been proposed: regularized LDA, which shrinks the sample covariance toward the identity \citep{friedman1989regularized}; optimal shrinkage estimation \citep{ledoit2004well}; sparse LDA with sparsity-inducing penalties \citep{clemmensen2011sparse,witten2011penalized}; and various factorization-based methods \citep{pourahmadi1999joint,bickel2008covariance,rothman2010sparse}. These methods improve covariance estimation quality but do not directly optimize the precision matrix for classification performance.

A key observation motivating our work is that the cross-entropy loss of the LDA discriminant function is convex in the precision matrix $\Sigma^{-1}$ (see Theorem~\ref{thm:convex}). This means that optimizing $\Sigma^{-1}$ directly for classification has a well-behaved loss landscape, unlike generic non-convex optimization problems. Building on this, we propose LDA with gradient optimization (LDA-GO), which parametrizes $\Sigma^{-1} = LL^\top + \sigma^2 I$ with $L \in \mathbb{R}^{p \times d}$ and optimizes $L$ via gradient-based optimization. When $p > d$, the low-rank factorization reduces the number of free parameters from $p(p+1)/2$ to $pd$, acting as an implicit form of regularization. The algorithm automatically selects between a Gaussian negative log-likelihood and a cross-entropy loss based on data-driven structural diagnostics, adapting to the signal structure without user tuning. Our algorithm provides a scalable direct gradient computation that works entirely in the $d$-dimensional embedded space, achieving a per-iteration cost of $\mathcal{O}(npd)$.

Our theoretical analysis establishing convexity of the loss in $\Sigma^{-1}$, global optimality of local minima under sufficient rank, Bayes optimality, and an excess risk bound of $\mathcal{O}(pd/n)$. Our experiments compare LDA-GO against popular baselines, including classical LDA, Ledoit-Wolf LDA, logistic regression, Naive Bayes, random forest, and neural network, across 20 simulations and 5 real data. They demonstrate that LDA-GO wins 17/20 simulation settings and is consistently among the top two methods on real data.

In the following, we first discuss the necessary background, followed by introducing various components and the full algorithm of LDA-GO. Then we present theoretical properties, simulations, and real data experiments. Theorem proofs are provided in the appendix, and the Python code implementing LDA-GO and all experiments are provided in the supplementary materials.

\section{Background}

\subsection{Linear Discriminant Analysis}

Linear discriminant analysis (LDA) assumes each class-conditional distribution is Gaussian with shared covariance matrix $\Sigma \in \mathbb{R}^{p \times p}$ and class-specific means $\mu_k \in \mathbb{R}^{p}$.  Under Bayes' rule, the LDA discriminant function is:
\begin{align*}
\delta_k(x) = x^\top \Sigma^{-1} \mu_k - \frac{1}{2} \mu_k^\top \Sigma^{-1} \mu_k + \log \pi_k,
\end{align*}
where $\pi_k$ is the prior probability of class $k$.

Given training data $\{(x_i, y_i)\}_{i=1}^n$ with $x_i \in \mathbb{R}^p$ and $y_i \in \{1, \dots, K\}$, the parameters are estimated as:
\begin{align*}
\hat{\pi}_k &= \frac{n_k}{n}, \qquad
\hat{\mu}_k = \frac{1}{n_k} \sum_{i: y_i = k} x_i, \\
\hat{\Sigma} &= \frac{1}{n - K} \sum_{k=1}^K \sum_{i: y_i = k} (x_i - \hat{\mu}_k)(x_i - \hat{\mu}_k)^\top,
\end{align*}
where $n_k$ is the number of training samples in class $k$. The predicted label is $\hat{y} = \arg\max_k \delta_k(x)$.

\subsection{Related Methods}

The sample covariance $\hat{\Sigma}$ becomes ill-conditioned or singular when $p \approx n$ or $p > n$. Popular existing approaches addressing this:

\textbf{Shrinkage LDA.}  Regularized LDA \citep{friedman1989regularized} replaces $\hat{\Sigma}$ with $(1 - \gamma) \hat{\Sigma} + \gamma I$ for a tuning parameter $\gamma \in (0,1)$.  The Ledoit-Wolf estimator \citep{ledoit2004well} selects $\gamma$ to minimize the expected Frobenius loss.  These methods improve covariance estimation quality but do not optimize the precision matrix for classification performance directly.

\textbf{Logistic regression.}  LDA and logistic regression both yield a linear decision boundary, but they differ in parametrization: LDA constrains the $K$ class-specific weight vectors to share a common precision matrix ($w_k = \Sigma^{-1}\mu_k$), while logistic regression learns $K$ independent weight vectors.  When the Gaussian model is correct, LDA can leverage the shared structure; when it is wrong, logistic regression can be more flexible.

\textbf{Gaussian Naive Bayes.}  Naive Bayes restricts the covariance to be diagonal, estimating each feature's variance independently per class.  This reduces the number of parameters from $p(p+1)/2$ to $Kp$, which can be advantageous when $p \gg n$, but the diagonal assumption discards all covariance information.

\textbf{Ensemble and neural methods.}  Random forests \citep{Breiman2001} and neural networks \citep{goodfellow2016deep} are widely used nonparametric classifiers. While they do not tackle the LDA covariance issue, they are designed to capture nonlinear decision boundaries by their own unique algorithm design. Specifically, random forests identify proper dimension splits and build multiple trees, while standard neural network minimizes a cross-entropy loss function by gradient descent.

\section{Method}

Given training data $\{(x_i, y_i)\}_{i=1}^n$, LDA-GO estimates class means and priors using sample-based methods, then learns a precision matrix $\Sigma^{-1}$ via gradient optimization.  The method has four stages: within-class standardization, precision parametrization, structural diagnostics for automatic loss selection, and gradient-based training.

\subsection{Within-Class Standardization}

Standardization is a common technique that ensures stable gradient dynamics when features have heterogeneous scales. Let $R_{ik} = x_i - \hat{\mu}_{y_i}$ denote the within-class residuals.  We compute $\bar{x} = \frac{1}{n}\sum_{i} x_i$ (the overall mean) and $s_j = \text{std}(\{R_{ij}\}_{i=1}^n)$ (the marginal within-class standard deviation for feature $j$).  The transformed data and means are:
\begin{align*}
\tilde{x}_i = \frac{x_i - \bar{x}}{s}, \qquad \tilde{\mu}_k = \frac{\hat{\mu}_k - \bar{x}}{s},
\end{align*}
where division is elementwise.  All subsequent optimization operates on $(\tilde{x}_i, \tilde{\mu}_k)$.  At prediction time, the same transformation is applied to new data.  This internal standardization ensures stable optimization regardless of the original feature scales.

\subsection{Precision Parametrization}

We then parametrize the precision matrix as:
\begin{align}\label{eq:factor}
\Sigma^{-1} = LL^\top + \sigma^2 I, \qquad L \in \mathbb{R}^{p \times d},
\end{align}
where $d = \min(20, p)$ is the rank parameter and $\sigma^2 \geq 0$ is a scalar.  The low-rank component $LL^\top$ captures the principal discriminative directions, while the diagonal $\sigma^2 I$ accounts for isotropic residual variance.  The factorization ensures that $\Sigma^{-1}$ is symmetric positive semi-definite throughout optimization.  Unlike shrinkage estimators that target the true covariance, LDA-GO optimizes the precision matrix directly for classification accuracy, a discriminative approach to covariance learning.

The embedded discriminant function is:
\begin{align*}
\delta_k(x_i) = z_i^\top w_k - \tfrac{1}{2}\|w_k\|^2 + \sigma^2\!\left(\tilde{x}_i^\top \tilde{\mu}_k - \tfrac{1}{2}\|\tilde{\mu}_k\|^2\right) + \log \hat{\pi}_k,
\end{align*}
where $z_i = L^\top \tilde{x}_i \in \mathbb{R}^d$ and $w_k = L^\top \tilde{\mu}_k \in \mathbb{R}^d$ are the embedded data and means.

\subsection{Structural Diagnostics and Loss Selection}
\label{sec:diagnostics}

LDA-GO automatically selects between two loss functions based on data-driven diagnostics computed from the standardized data:

\textbf{Signal sparsity} $r$.  Let $b_j = \sum_{k} \hat{\pi}_k (\tilde{\mu}_{kj} - \bar{\tilde{x}}_j)^2$ be the between-class variance along feature $j$.  The effective signal dimensionality is $D_{\text{eff}} = (\sum_j b_j)^2 / \sum_j b_j^2$, and the sparsity ratio is $r = D_{\text{eff}} / p$.  When $r$ is small, the discriminative signal is concentrated in few features.

\textbf{Kurtosis} $\kappa$.  Let $\kappa = \frac{1}{p}\sum_{j=1}^p |e_j - 3|$, where $e_j$ is the marginal kurtosis of the standardized within-class residuals along feature $j$.  Large $\kappa$ indicates departure from Gaussianity.

\textbf{Selection rule.}  If $r < 0.10$ or $\kappa > 10$, use cross-entropy (CE).  Otherwise, use negative log-likelihood (NLL).  Both diagnostics are $\mathcal{O}(np)$.

\begin{figure}[t]
\centering
\includegraphics[width=\columnwidth]{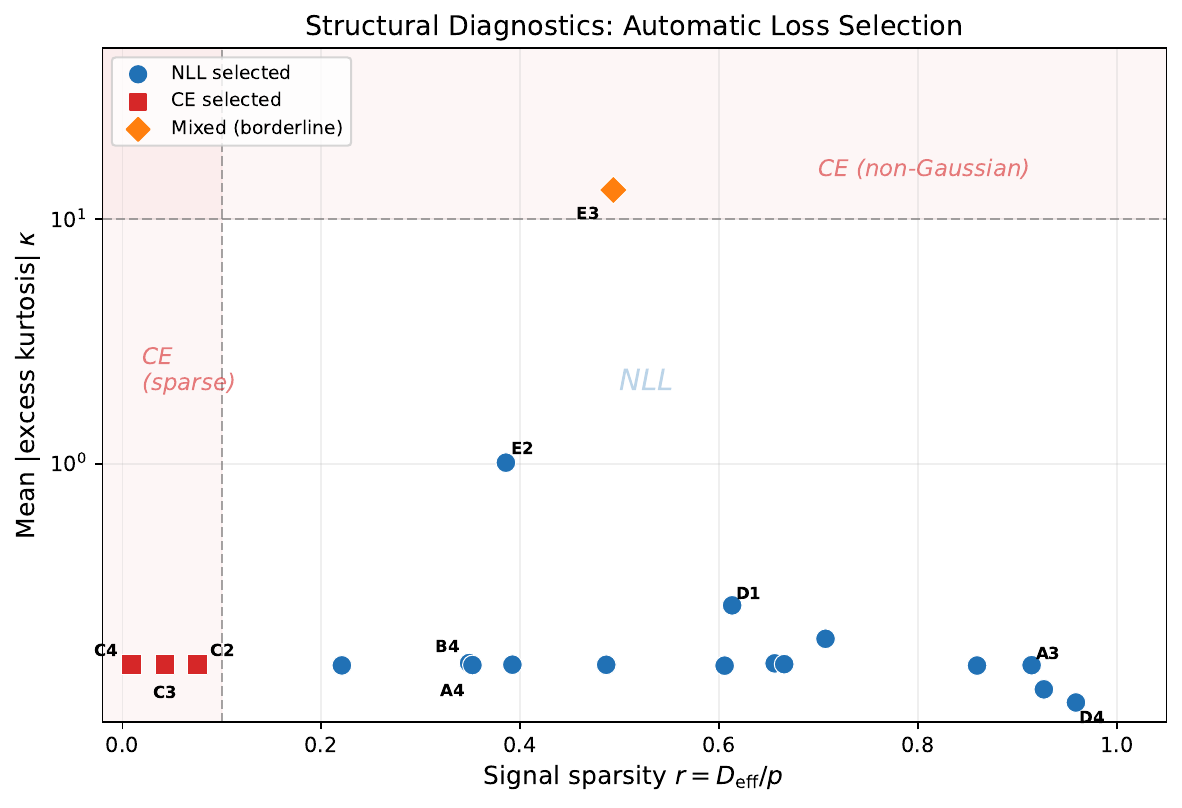}
\caption{Structural diagnostics for automatic loss selection across all 20 simulations (See Section~\ref{sec:sim}).  Each point represents one simulation's median $(r, \kappa)$ values.  Dashed lines show the decision boundaries $r = 0.10$ and $\kappa = 10$.  Blue circles: NLL selected; red squares: CE selected; orange diamond: borderline (mixed across replicates).}
\label{fig:diagnostics}
\end{figure}

\emph{Rationale.}  When the signal is dense and data are near-Gaussian, the NLL directly exploits the generative model.  When the signal is sparse or data are non-Gaussian, CE is preferable because it is a purely discriminative objective that does not commit to the Gaussian likelihood.  Moreover, the CE path uses plain gradient descent, which benefits from implicit regularization toward low-rank solutions in the overparameterized regime \citep{gunasekar2017implicit,li2018algorithmic}, while the NLL path uses Adam for faster convergence when the Gaussian model is well-specified.  Figure~\ref{fig:diagnostics} visualizes the diagnostic values $(r, \kappa)$ for all 20 simulation settings, showing clean separation between the NLL and CE regions.

\subsection{NLL Path}

Under the Gaussian model, the negative log-likelihood for the within-class residuals with covariance $\Sigma = (LL^\top + \sigma^2 I)^{-1}$ is minimized by choosing $L$ so that $LL^\top$ captures the structure of $\hat{\Sigma}_w^{-1}$ beyond the isotropic component.  The scalar $\sigma^2$ is set using Oracle Approximating Shrinkage \citep[OAS,][]{chen2010shrinkage}:  given the OAS shrinkage parameter $\alpha$ and trace $\mu = \text{tr}(\hat{\Sigma}_w)/p$, we set $\sigma^2 = 1/(\alpha \mu)$, bounded below by a small $\epsilon > 0$ for numerical stability.

The gradient of the NLL with respect to $L$ is:
\begin{align}\label{eq:grad_nll}
\frac{\partial \mathcal{L}_{\text{NLL}}}{\partial L} = \hat{\Sigma}_{w,\alpha} L - L \left(\sigma^2 I_d + L^\top L\right)^{-1},
\end{align}
where $\hat{\Sigma}_{w,\alpha} = (1 - \alpha)\hat{\Sigma}_w + \alpha \mu I$ is the OAS-shrunk within-class covariance.  Optimization uses Adam \citep{kingma2014adam} with $\eta = 0.01$ for up to 500 iterations.

\subsection{CE Path}

The cross-entropy loss is:
\begin{align}\label{eq:loss}
\mathcal{L}_{\text{CE}}(L) = -\frac{1}{n} \sum_{i=1}^n \sum_{k=1}^K Y_{ik} \log P_{ik},
\end{align}
where $Y_{ik} = \mathbf{1}[y_i = k]$ and $P_{ik} = \text{softmax}_k(\delta(x_i))$.  In this path, $\sigma^2 = 0$ (no isotropic component).

The gradient of the CE loss can be expressed without forming any $p \times p$ matrix.  Let the residual matrix $\mathcal{R} \in \mathbb{R}^{n \times K}$ have entries $\mathcal{R}_{ik} = Y_{ik} - P_{ik}$.  Then:
\begin{align}\label{eq:grad_direct}
\frac{\partial \mathcal{L}_{\text{CE}}}{\partial L} = -\frac{1}{n}\!\left( \tilde{X}^\top \mathcal{R}\, W + \tilde{M}^\top (\mathcal{R}^\top Z) - \tilde{M}^\top \text{diag}(\mathbf{1}^\top \mathcal{R})\, W \right),
\end{align}
where $\tilde{X} \in \mathbb{R}^{n \times p}$ is the standardized data matrix, $\tilde{M} \in \mathbb{R}^{K \times p}$ is the standardized class means matrix, $Z = \tilde{X}L \in \mathbb{R}^{n \times d}$, and $W = \tilde{M}L \in \mathbb{R}^{K \times d}$.  The total cost is $\mathcal{O}(npd + nKp)$ per iteration.  Optimization uses gradient descent with $\eta = 0.01$ for up to 500 iterations.

\subsection{Summary and Algorithm}

LDA-GO proceeds in four stages:
\begin{itemize}
  \item \textbf{Standardize.}  Estimate class means and priors; standardize features using within-class residual statistics.
  \item \textbf{Parametrize.}  Set $\Sigma^{-1} = LL^\top + \sigma^2 I$ with $L \in \mathbb{R}^{p \times d}$, $d = \min(20, p)$.
  \item \textbf{Auto Loss Selection.}  Compute signal sparsity $r$ and excess kurtosis $\kappa$ in $\mathcal{O}(np)$ time; route to CE if $r < 0.10$ or $\kappa > 10$, otherwise NLL.
  \item \textbf{Optimize.}  Initialize $L^{(0)} = I_{p \times d} + \epsilon$, $\epsilon_{ij} \sim \mathcal{N}(0, 0.01^2)$, and run up to $T = 500$ gradient iterations (Adam for NLL, gradient descent for CE, both with $\eta = 0.01$), stopping early when $\|\nabla \mathcal{L}\| < 10^{-4}$.
\end{itemize}
Both loss functions are convex in $\Sigma^{-1}$ (Theorem~\ref{thm:convex}), and the low-rank factorization keeps the per-iteration cost at $\mathcal{O}(npd)$, comparable to logistic regression, without forming any $p \times p$ intermediate matrix.  The full procedure is given in Algorithm~\ref{alg:ldago}.

\begin{algorithm}[t]
\caption{LDA-GO}
\label{alg:ldago}
\begin{algorithmic}[1]
\Require Training data $\{(x_i, y_i)\}_{i=1}^n$, rank $d = \min(20, p)$, max iterations $T = 500$
\Ensure Precision factor $L^*$, scalar $\sigma^{2*}$
\State Estimate class means $\hat{\mu}_k$, priors $\hat{\pi}_k$
\State Compute within-class standardization: $\bar{x}$, $s$ \Comment{$\mathcal{O}(np)$}
\State Standardize: $\tilde{x}_i \leftarrow (x_i - \bar{x})/s$, $\tilde{\mu}_k \leftarrow (\hat{\mu}_k - \bar{x})/s$
\State Compute diagnostics: sparsity $r$, kurtosis $\kappa$ \Comment{$\mathcal{O}(np)$}
\If{$r < 0.10$ \textbf{or} $\kappa > 10$}
  \State \textbf{CE path:} $\sigma^2 \leftarrow 0$
  \State Initialize $L \leftarrow I_{p \times d} + \epsilon$
  \For{$t = 1, \dots, T$}
    \State Compute CE gradient~\eqref{eq:grad_direct}; $L \leftarrow L - 0.01 \cdot \nabla \mathcal{L}_{\text{CE}}$
    \State \textbf{if} $\|\nabla \mathcal{L}_{\text{CE}}\| < 10^{-4}$ \textbf{then break}
  \EndFor
\Else
  \State \textbf{NLL path:} Compute OAS shrinkage $(\alpha, \mu)$; set $\sigma^2 \leftarrow 1/(\alpha\mu)$
  \State Initialize $L \leftarrow I_{p \times d} + \epsilon$
  \For{$t = 1, \dots, T$}
    \State Compute NLL gradient~\eqref{eq:grad_nll}; update $L$ via Adam ($\eta = 0.01$)
    \State \textbf{if} $\|\nabla \mathcal{L}_{\text{NLL}}\| < 10^{-4}$ \textbf{then break}
  \EndFor
\EndIf
\State \Return $L^* \leftarrow L$, $\sigma^{2*} \leftarrow \sigma^2$
\end{algorithmic}
\end{algorithm}

\textbf{Complexity.}  Preprocessing (standardization + diagnostics) costs $\mathcal{O}(np)$.  Each gradient iteration costs $\mathcal{O}(npd + nKp)$ for the CE path and $\mathcal{O}(npd + d^3)$ for the NLL path; both reduce to $\mathcal{O}(npd)$ when $d$ and $K$ are small constants.  The total training cost is $\mathcal{O}(T \cdot npd)$.  With $d = 20$ and $T = 500$, this amounts to $10{,}000 \cdot np$ operations, comparable to logistic regression and far below the $\mathcal{O}(np^2 + p^3)$ cost of classical LDA.

\textbf{Connection to logistic regression.}  LDA-GO and logistic regression both produce linear decision boundaries, but they differ in parametrization: LDA-GO constrains the $K$ class-specific weight vectors to share a common precision matrix structure ($w_k = \Sigma^{-1}\mu_k$), while logistic regression learns $K$ independent weight vectors.  When the shared-covariance assumption holds, LDA-GO's parametrization reduces the effective number of free parameters and can generalize better; when it does not, logistic regression is more flexible.

\textbf{Choice of rank $d$.}  We use $d = \min(20, p)$ throughout.  Empirically, classification performance is stable for $d \geq 20$, with larger values slightly increasing computation but not substantially improving accuracy.  When the discriminative subspace has very low intrinsic dimension, a smaller $d$ suffices.

\textbf{When to use LDA-GO.}  LDA-GO is most effective when: (i) $p$ is large relative to $n$, making classical covariance estimation unreliable; (ii) the discriminative signal is sparse or lies in a low-dimensional subspace; and (iii) a linear decision boundary is appropriate.  In the low-dimensional regime ($p \ll n$), classical LDA or logistic regression are sufficiently effective.

\section{Theoretical Analysis}

In this section, we establish four theoretical properties of LDA-GO. Throughout, we write $\mathbb{S}^p_+$ for the cone of $p \times p$ symmetric positive semi-definite matrices.  We use $\mathcal{L}(\Sigma^{-1})$ to denote either loss function (CE or NLL) viewed as a function of the precision matrix, and $\mathcal{L}_n$, $\mathcal{L}_\infty$ for its empirical and population versions. All proofs are deferred to the Appendix, which apply various techniques from \citet{boyd2004convex,burer2003nonlinear,bhojanapalli2016dropping, van2000asymptotic}.

\begin{theorem}[Convexity of the LDA-GO Objective]\label{thm:convex}
Both loss functions used by LDA-GO, i.e., the cross-entropy loss $\mathcal{L}_{\text{CE}}(\Sigma^{-1})$ defined in~\eqref{eq:loss} and the Gaussian negative log-likelihood $\mathcal{L}_{\text{NLL}}(\Sigma^{-1})$, are convex in $\Sigma^{-1} \in \mathbb{S}^p_+$.
\end{theorem}

Theorem~\ref{thm:convex} implies that, regardless of which loss LDA-GO selects, the optimization landscape over $\Sigma^{-1}$ has no spurious local minima. However, LDA-GO optimizes a factored objective $\mathcal{L}(LL^\top + \sigma^2 I)$ with $\sigma^2$ fixed, which is non-convex in $L$. The following theorem shows that this non-convexity is benign when $d$ is sufficiently large, where a local minimum is also global minimum.

\begin{theorem}[Global Optimality of Local Minima]\label{thm:global}
Let $\Sigma^{*-1} = \arg\min_{\Sigma^{-1} \succeq \sigma^2 I} \mathcal{L}(\Sigma^{-1})$ denote the global minimizer over the feasible set, with $r^* = \text{rank}(\Sigma^{*-1} - \sigma^2 I)$, where $\sigma^2$ is the fixed scalar in~\eqref{eq:factor}. If the factorization rank satisfies $d \geq r^*$, then every local minimum $L^*$ of $\mathcal{L}(LL^\top + \sigma^2 I)$ is a global minimum: $\mathcal{L}(L^* L^{*\top} + \sigma^2 I) = \mathcal{L}(\Sigma^{*-1})$.
\end{theorem}

Our next theorems show that the LDA-GO can still converge to Bayes error just like classical LDA, and the finite-sample excess risk is linear.

\begin{theorem}[Consistency]\label{thm:consistency}
Suppose the true data-generating distribution satisfies $X \mid Y = k \sim \mathcal{N}(\mu_k^0, \Sigma_0)$ with $\Sigma_0 \succ 0$, and the class means $\hat{\mu}_k$ and priors $\hat{\pi}_k$ are consistent estimators of $\mu_k^0$ and $\pi_k^0$, respectively. Let $d = p$ and let $\hat{L}_n$ denote the global minimizer of $\mathcal{L}_n(LL^\top + \sigma^2 I)$ over the training data. Then as $n \to \infty$, the classification error of LDA-GO converges to the Bayes error rate.
\end{theorem}.

\begin{theorem}[Excess Risk Bound]\label{thm:risk}
Under the conditions of Theorem~\ref{thm:consistency}, suppose additionally that the class means $\mu_k^0$ are known and that $\Sigma_0^{-1} - \sigma^2 I$ has rank at most $d$. Then the excess misclassification risk of LDA-GO satisfies:
\[
\text{Risk}(\hat{L}_n \hat{L}_n^\top + \sigma^2 I) - \text{Risk}(\Sigma_0^{-1}) = \mathcal{O}_P\!\left(\frac{pd}{n}\right).
\]
\end{theorem}

Note that the cross-entropy loss depends on $\Sigma^{-1}$ only through the posterior class probabilities, which for $K$ classes are determined by $\Sigma^{-1}(\mu_k^0 - \mu_j^0)$ for all class pairs $(k,j)$. When $K - 1 < p$, different precision matrices can yield the same posteriors. However, the convergence of class posteriors, and hence classification error, is unaffected

\section{Simulations}
\label{sec:sim}

We conduct a comprehensive simulation study to evaluate LDA-GO across 20 settings organized into 5 categories, each isolating one factor. Unless otherwise noted, we set $n_{\text{train}} = 200$, $n_{\text{test}} = 10{,}000$, and $K = 2$ with equal priors. 

\subsection{Simulation Settings}

The 20 simulations are summarized in Table~\ref{tab:simulation_summary}.

\textbf{Category A: Covariance structure} ($p = 200$, dense calibrated $\Delta\mu$). Varies the covariance from identity through diagonal and strong Toeplitz to random dense:
A1: $\Sigma = I$;
A2: $\Sigma = \text{diag}(1, 4, 9, \dots, 100, 1, \dots, 1)$ (first 10 dims have increasing variance);
A3: $\Sigma = \text{Toeplitz}(0.9)$;
A4: $\Sigma = AA^\top/p$ with $A$ having i.i.d.\ $\mathcal{N}(0,1)$ entries, which is a trivial no-signal setting.

\textbf{Category B: Dimensionality} ($\Sigma = I$, dense calibrated $\Delta\mu$). Varies $p$ from the classical regime through severely overparameterized:
B1: $p = 20$;
B2: $p = 100$;
B3: $p = 500$;
B4: $p = 1000$.

\textbf{Category C: Signal sparsity} ($p = 500$, $\Sigma = I$). Varies the number of active signal dimensions from moderate to a single one:
C1: 50/500 active;
C2: 10/500 active;
C3: 5/500 active;
C4: 1/500 active (needle-in-haystack).

\textbf{Category D: Sample size} ($p = 200$, $\Sigma = \text{Toeplitz}(0.8)$, dense calibrated $\Delta\mu$). Varies $n$ from severely undersampled to ample:
D1: $n = 50$ ($p/n = 4$);
D2: $n = 100$ ($p/n = 2$);
D3: $n = 500$ ($p/n = 0.4$);
D4: $n = 1000$ ($p/n = 0.2$).

\textbf{Category E: Model violations} ($p = 200$). Tests robustness when the shared-Gaussian assumption fails:
E1: Gaussian with shared $\Sigma = \text{Toeplitz}(0.5)$ (control);
E2: class-dependent scale ($\Sigma_0 = I$, $\Sigma_1 = 4I$);
E3: heavy tails (multivariate $t$ with $\text{df} = 3$, shared $\Sigma = \text{Toeplitz}(0.5)$);
E4: mixture of Gaussians (each class is a 2-component mixture with shared $\Sigma = \text{Toeplitz}(0.5)$).

\begin{table*}[htbp]
\centering
\caption{Summary of the 20 simulation settings in 5 categories. Each category varies one factor while holding others fixed. The mean difference $\Delta\mu$ is calibrated so that the Bayes error $\approx 6.7\%$ when the model is correct.}
\begin{tabular}{|c|c|l|}
\hline
\textbf{ID} & \textbf{Category} & \textbf{Description} \\ \hline
A1 & \multirow{4}{*}{Covariance} & $\Sigma = I$, $p = 200$ \\ \cline{1-1}\cline{3-3}
A2 & & $\Sigma = \text{diag}(1,4,\dots,100,1,\dots)$, $p = 200$ \\ \cline{1-1}\cline{3-3}
A3 & & $\Sigma = \text{Toeplitz}(0.9)$, $p = 200$ \\ \cline{1-1}\cline{3-3}
A4 & & $\Sigma = AA^\top/p$ (random dense), $p = 200$ \\ \hline
B1 & \multirow{4}{*}{Dimension} & $p = 20$, $\Sigma = I$ \\ \cline{1-1}\cline{3-3}
B2 & & $p = 100$, $\Sigma = I$ \\ \cline{1-1}\cline{3-3}
B3 & & $p = 500$, $\Sigma = I$ \\ \cline{1-1}\cline{3-3}
B4 & & $p = 1000$, $\Sigma = I$ \\ \hline
C1 & \multirow{4}{*}{Sparsity} & 50/500 dims active, $\Sigma = I$ \\ \cline{1-1}\cline{3-3}
C2 & & 10/500 dims active, $\Sigma = I$ \\ \cline{1-1}\cline{3-3}
C3 & & 5/500 dims active, $\Sigma = I$ \\ \cline{1-1}\cline{3-3}
C4 & & 1/500 dims active, $\Sigma = I$ \\ \hline
D1 & \multirow{4}{*}{Sample size} & $n = 50$ ($p/n = 4$), $\Sigma = \text{Toep}(0.8)$ \\ \cline{1-1}\cline{3-3}
D2 & & $n = 100$ ($p/n = 2$), $\Sigma = \text{Toep}(0.8)$ \\ \cline{1-1}\cline{3-3}
D3 & & $n = 500$ ($p/n = 0.4$), $\Sigma = \text{Toep}(0.8)$ \\ \cline{1-1}\cline{3-3}
D4 & & $n = 1000$ ($p/n = 0.2$), $\Sigma = \text{Toep}(0.8)$ \\ \hline
E1 & \multirow{4}{*}{Violations} & Gaussian shared $\Sigma = \text{Toeplitz}(0.5)$ (control) \\ \cline{1-1}\cline{3-3}
E2 & & Class-dependent scale ($\Sigma_0 = I$, $\Sigma_1 = 4I$) \\ \cline{1-1}\cline{3-3}
E3 & & Heavy tails ($t$, $\text{df}=3$) \\ \cline{1-1}\cline{3-3}
E4 & & Mixture of Gaussians \\ \hline
\end{tabular}
\label{tab:simulation_summary}
\end{table*}

\subsection{Comparison Methods}

We compare LDA-GO against six methods:
\begin{itemize}
\item \textbf{LDA:} Classical LDA via singular value decomposition.
\item \textbf{LW-LDA:} LDA with Ledoit-Wolf optimal shrinkage \citep{ledoit2004well}.
\item \textbf{LogReg:} $L_2$-regularized logistic regression ($C = 1$, LBFGS solver).
\item \textbf{NB:} Gaussian Naive Bayes.
\item \textbf{RF:} Random forest with 100 trees.
\item \textbf{NN:} Two-layer neural network (64 hidden units, early stopping).
\end{itemize}
All baselines use default parameters from \texttt{scikit-learn} \citep{pedregosa2011scikit}. We intentionally exclude slow methods (e.g., sparse LDA, graphical lasso) to focus on methods that scale to $p > 1000$. For each simulation, we conduct 20 Monte Carlo replicates and report the mean classification error on a held-out test set of $n_{\text{test}} = 10{,}000$.

\subsection{Classification Results}

Table~\ref{tab:simulation_results} reports error rates across all 20 settings. LDA-GO wins or ties for the best error rate in 17 of 20 simulations.

\begin{table*}[htbp]
\centering
\caption{Average classification error rates (\%) across 20 simulations, 20 Monte Carlo replicates. Bold indicates the best method. LDA-GO wins 17/20 settings.}
\renewcommand{\arraystretch}{1.1}
\begin{tabular}{c|c|c|c|c|c|c|c|c}
\hline \hline
ID & $p$ & LDA-GO & LDA & LW-LDA & NB & LogReg & RF & NN \\
\hline \hline
\multicolumn{9}{l}{\textit{Category A: Covariance Structure} ($p = 200$)} \\ \hline
A1 & 200 & $\mathbf{11.0}$ & $43.5$ & $\mathbf{11.0}$ & $14.4$ & $14.1$ & $24.5$ & $40.4$ \\
A2 & 200 & $\mathbf{10.6}$ & $42.0$ & $\mathbf{10.6}$ & $13.9$ & $14.5$ & $24.2$ & $43.5$ \\
A3 & 200 & $\mathbf{7.3}$ & $44.1$ & $10.1$ & $7.4$ & $9.9$ & $9.3$ & $23.3$ \\
A4 & 200 & $49.6$ & $47.9$ & $49.6$ & $49.7$ & $49.3$ & $49.8$ & $49.9$ \\
\hline
\multicolumn{9}{l}{\textit{Category B: Dimensionality} ($\Sigma = I$)} \\ \hline
B1 & 20 & $\mathbf{7.3}$ & $8.0$ & $\mathbf{7.3}$ & $7.7$ & $8.2$ & $10.9$ & $25.0$ \\
B2 & 100 & $\mathbf{9.0}$ & $16.6$ & $\mathbf{9.0}$ & $11.1$ & $12.3$ & $18.7$ & $23.0$ \\
B3 & 500 & $\mathbf{15.5}$ & $31.0$ & $\mathbf{15.5}$ & $20.6$ & $17.6$ & $33.7$ & $33.4$ \\
B4 & 1000 & $\mathbf{20.7}$ & $37.9$ & $\mathbf{20.6}$ & $26.6$ & $21.9$ & $39.6$ & $39.2$ \\
\hline
\multicolumn{9}{l}{\textit{Category C: Signal Sparsity} ($p = 500$, $\Sigma = I$)} \\ \hline
C1 & 500 & $\mathbf{15.2}$ & $30.1$ & $\mathbf{15.2}$ & $20.3$ & $17.4$ & $26.0$ & $35.3$ \\
C2 & 500 & $\mathbf{8.9}$ & $30.5$ & $15.5$ & $20.6$ & $17.9$ & $17.0$ & $42.4$ \\
C3 & 500 & $\mathbf{8.9}$ & $31.3$ & $15.7$ & $20.9$ & $17.6$ & $14.3$ & $46.1$ \\
C4 & 500 & $\mathbf{8.8}$ & $30.8$ & $15.4$ & $21.0$ & $17.5$ & $12.9$ & $50.3$ \\
\hline
\multicolumn{9}{l}{\textit{Category D: Sample Size} ($p = 200$, $\Sigma = \text{Toeplitz}(0.8)$)} \\ \hline
D1 & 200 & $\mathbf{8.0}$ & $12.8$ & $11.1$ & $8.7$ & $10.3$ & $13.1$ & $42.8$ \\
D2 & 200 & $\mathbf{7.6}$ & $15.3$ & $10.3$ & $7.9$ & $9.7$ & $12.0$ & $35.8$ \\
D3 & 200 & $\mathbf{7.0}$ & $14.2$ & $8.7$ & $\mathbf{7.0}$ & $9.5$ & $9.0$ & $13.8$ \\
D4 & 200 & $\mathbf{6.9}$ & $10.0$ & $8.2$ & $\mathbf{6.9}$ & $10.1$ & $8.5$ & $10.2$ \\
\hline
\multicolumn{9}{l}{\textit{Category E: Model Violations} ($p = 200$)} \\ \hline
E1 & 200 & $\mathbf{7.7}$ & $44.8$ & $8.4$ & $8.2$ & $10.4$ & $14.0$ & $36.0$ \\
E2 & 200 & $23.5$ & $44.9$ & $23.5$ & $\mathbf{0.0}$ & $25.9$ & $0.6$ & $46.6$ \\
E3 & 200 & $14.7$ & $44.4$ & $\mathbf{13.5}$ & $36.6$ & $16.2$ & $19.4$ & $38.3$ \\
E4 & 200 & $\mathbf{9.5}$ & $44.2$ & $10.4$ & $9.9$ & $12.9$ & $15.5$ & $27.3$ \\
\hline \hline
\end{tabular}
\label{tab:simulation_results}
\end{table*}

\textbf{Categories A--B.} LDA-GO wins or ties the best performance in all 7 non-degenerate settings. In Categories A and B it is closely matched with LW-LDA, both outperforming classical LDA with wider margin as $p/n$ grows. The only exception is A4, where all methods perform near chance due to no-signal.

\begin{figure}[t]
\centering
\includegraphics[width=\columnwidth]{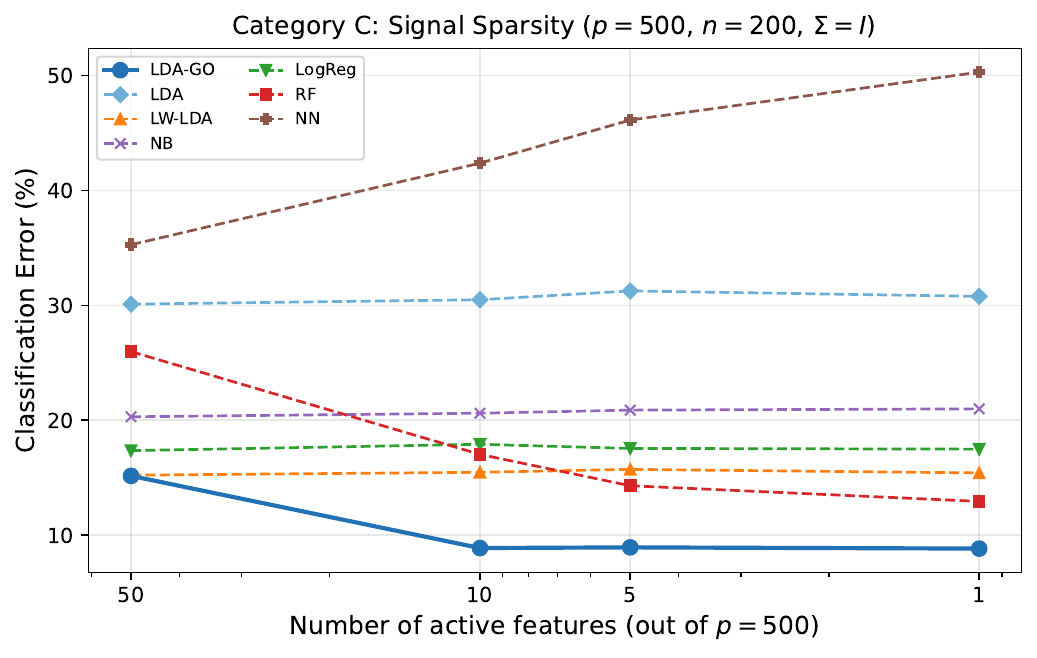}
\caption{Classification error versus signal sparsity (Category C: $p = 500$, $n = 200$, $\Sigma = I$).  As the number of active features decreases from 50 to 1, LDA-GO maintains low error while other methods remain substantially higher.}
\label{fig:sparsity}
\end{figure}

\textbf{Category C (signal sparsity).} LDA-GO dominates all 4 settings, and the advantage is most dramatic as signal becomes sparser: in C4 (1/500 active), LDA-GO achieves 8.8\% versus 12.9\% for RF and 15.4\% for LW-LDA (Figure~\ref{fig:sparsity}). The low-rank factorization naturally focuses on the discriminative subspace, avoiding the curse of estimating a $500 \times 500$ precision matrix from 200 samples.

\textbf{Category D--E.} In Category D, LDA-GO is best or tied at every sample size from $n = 50$ to $n = 1000$. In Category E, LDA-GO wins 2/4 model violation settings (E1, E4), demonstrating reasonable robustness. 

\section{Real Data Experiments}
\label{sec:real}

We evaluate LDA-GO on five benchmark datasets spanning gene expression, voice rehabilitation, text classification, and face recognition:
\begin{enumerate}
\item \textbf{Leukemia} ($n = 72$, $p = 7{,}129$, $K = 2$): mRNA expression profiles of acute leukemia patients, classifying ALL vs.\ AML subtypes \citep{golub1999molecular}.
\item \textbf{Lung} ($n = 203$, $p = 12{,}600$, $K = 5$): gene expression profiles of human lung carcinomas across five histological subtypes \citep{bhattacharjee2001lung}.
\item \textbf{Olivetti Faces} ($n = 100$, $p = 4{,}096$, $K = 10$): grayscale face images of a 10-subject subset of the ORL face database under varying lighting and expression \citep{SamariaHarter1994}.
\item \textbf{LSVT} ($n = 126$, $p = 307$, $K = 2$): voice features extracted from sustained vowel phonations, classifying acceptable vs.\ unacceptable speech rehabilitation outcomes in Parkinson's disease patients \citep{tsanas2014lsvt}.
\item \textbf{CNAE-9} ($n = 1{,}080$, $p = 856$, $K = 9$): term-frequency vectors from free-text business descriptions of Brazilian companies across 9 economic sectors \citep{ciarelli2009cnae}.
\end{enumerate}

For the three very-high-dimensional datasets (Leukemia, Lung, Olivetti) where $p \gg n$, we first apply PCA to the full dataset retaining 99\% of the total variance, reducing dimensionality to $p' \in [64, 169]$. Since PCA is unsupervised (it does not use class labels), this introduces no label leakage. The two moderate-dimensional datasets (LSVT, CNAE-9) are used without dimensionality reduction. Performance is assessed via 5-fold cross-validation repeated 5 times. Within each fold, features are standardized to zero mean and unit variance using training-set statistics. Results are reported in Table~\ref{tab:data_results}.

\begin{table*}[htbp]
\centering
\caption{Average classification error rates (\%) on five benchmark datasets, 5-fold CV $\times$ 5 repeats. For Leukemia, Lung, and Olivetti, PCA retaining 99\% variance is applied before classification; LSVT and CNAE-9 use the original features. Bold indicates the best method; $p'$ denotes the dimensionality used for classification.}
\renewcommand{\arraystretch}{1.1}
\begin{tabular}{l|cccc|ccccccc}
\hline \hline
Dataset & $n$ & $p$ & $p'$ & $K$ & LDA-GO & LDA & LW-LDA & NB & LogReg & RF & NN \\
\hline \hline
Leukemia  & 72    & 7{,}129  & 64  & 2  & $\mathbf{4.8}$ & $43.9$ & $12.8$ & $11.3$ & $6.2$  & $10.4$ & $50.6$ \\
Lung      & 203   & 12{,}600 & 169 & 5  & $\mathbf{5.1}$ & $66.5$ & $23.0$ & $22.7$ & $11.1$ & $20.3$ & $49.0$ \\
Olivetti  & 100   & 4{,}096  & 79  & 10 & $4.4$          & $52.8$ & $\mathbf{3.6}$ & $37.2$ & $4.2$  & $14.2$ & $94.2$ \\
LSVT      & 126   & 307      & 307 & 2  & $17.0$         & $29.3$ & $26.9$ & $56.0$ & $33.5$ & $\mathbf{16.3}$ & $41.9$ \\
CNAE-9    & 1{,}080 & 856    & 856 & 9  & $\mathbf{5.5}$          & $14.0$ & $21.4$ & $10.2$ & $\mathbf{5.5}$  & $7.7$  & $7.1$ \\
\hline \hline
\end{tabular}
\label{tab:data_results}
\end{table*}

LDA-GO achieves the lowest error rate on Leukemia and Lung. On Olivetti, LDA-GO  closely trails LW-LDA and ranks third. On LSVT, LDA-GO is very close to RF and ranks second. On CNAE-9, the largest dataset, LDA-GO is essentially tied with LogReg. Across all five datasets, LDA-GO is consistently the best or close to the best error, displaying clear advantages and robustness against all other methods.

\section{Conclusion}

We proposed LDA-GO, a gradient-based approach to learn the precision matrix in linear discriminant analysis. The method uses a low-rank-plus-diagonal precision parametrization, automatically selects between a Gaussian likelihood and a cross-entropy loss via structural diagnostics, and employs a scalable direct gradient that avoids forming any quadratic-sized intermediate matrix.  We established convexity of both losses in the precision matrix, global optimality of local minima for the factored problem, Bayes optimal classification property, and an excess risk bound that scales with the intrinsic rank rather than the ambient dimension. The simulations and real data experiments clearly demonstrate the advantages of the proposed method.

There are several possible future directions.  First, the framework could be extended to quadratic discriminant analysis by learning class-specific precision matrices, each with its own low-rank factorization.  Second, mini-batch gradient descent and online updates would enable scaling to very large sample sizes.  Third, a formal analysis of the automatic loss selection mechanism, characterizing when the Gaussian likelihood outperforms cross-entropy and vice versa, may provide extra insights.  Fourth, connections to neural network classifiers with shared linear output structure suggest potential hybrid architectures that combine the structural benefits of LDA-GO with deep feature extractors. Finally, extending the low-rank precision approach to multi-task and transfer learning settings, where related classification problems share covariance structure, is also an interesting avenue.

\bibliographystyle{apalike}
\bibliography{general, shen}

\newpage
\setcounter{figure}{0}
\renewcommand{\thealgorithm}{C\arabic{algorithm}}
\renewcommand{\thefigure}{E\arabic{figure}}
\renewcommand{\thesubsection}{\thesection.\arabic{subsection}}
\renewcommand{\thesubsubsection}{\thesubsection.\arabic{subsubsection}}

\appendix
\bigskip
\begin{center}
{\large\bf APPENDIX}
\end{center}

\section{Proofs}
\label{sec:proofs}

\subsection{Proof of Theorem~\ref{thm:convex}}
\begin{proof}
We show that each loss is a composition of a convex function with an affine mapping.

\emph{Cross-entropy.}  For each sample $x_i$ and class $k$, the discriminant function $\delta_k(x_i) = x_i^\top \Sigma^{-1} \hat{\mu}_k - \frac{1}{2}\hat{\mu}_k^\top \Sigma^{-1} \hat{\mu}_k + \log \hat{\pi}_k$ is affine in $\Sigma^{-1}$, since $x_i^\top \Sigma^{-1} \hat{\mu}_k = \operatorname{tr}(\Sigma^{-1} \hat{\mu}_k x_i^\top)$ and $\hat{\mu}_k^\top \Sigma^{-1} \hat{\mu}_k = \operatorname{tr}(\Sigma^{-1} \hat{\mu}_k \hat{\mu}_k^\top)$ are linear in $\Sigma^{-1}$. The per-sample cross-entropy $\ell_i(\delta) = -\delta_k(x_i) + \log \sum_{j} \exp(\delta_j(x_i))$ is convex in $\delta$ because $\log\!\sum \exp$ is convex \citep{boyd2004convex} and $-\delta_k$ is linear. Composing a convex function with an affine mapping preserves convexity, so $\ell_i(\Sigma^{-1})$ is convex, and $\mathcal{L}_{\text{CE}} = \frac{1}{n}\sum_i \ell_i$ is convex.

\emph{Negative log-likelihood.}  The Gaussian NLL for the within-class residuals is $\mathcal{L}_{\text{NLL}}(\Sigma^{-1}) = \tfrac{1}{2}\operatorname{tr}(\hat{\Sigma}_w \, \Sigma^{-1}) - \tfrac{1}{2}\log\det(\Sigma^{-1}) + \text{const}$.  The trace term is linear in $\Sigma^{-1}$, and $-\log\det$ is convex on $\mathbb{S}^p_+$ (see, e.g., \citet{boyd2004convex}), so $\mathcal{L}_{\text{NLL}}$ is convex.
\end{proof}

\subsection{Proof of Theorem~\ref{thm:global}}
\begin{proof}
We verify the conditions of the Burer-Monteiro framework \citep{burer2003nonlinear,bhojanapalli2016dropping} and apply it to our setting.

\emph{Reformulation.} For fixed $\sigma^2 \geq 0$, define $g(L) := \mathcal{L}(LL^\top + \sigma^2 I)$, where $L \in \mathbb{R}^{p \times d}$.  Because $\sigma^2 I$ is a constant shift, the mapping $L \mapsto LL^\top + \sigma^2 I$ traces out the set $\{M \in \mathbb{S}^p_+ : M \succeq \sigma^2 I,\; \text{rank}(M - \sigma^2 I) \leq d\}$.  Let $\Sigma^{*-1}$ denote the global minimizer of $\mathcal{L}$ over $\{\Sigma^{-1} \succeq \sigma^2 I\}$, with $r^* = \text{rank}(\Sigma^{*-1} - \sigma^2 I)$.  When $d \geq r^*$, the minimizer lies in this set, so $\Sigma^{*-1} - \sigma^2 I = L^*(L^*)^\top$ for some $L^*$.

\emph{Verification of conditions.} The function $h(Q) := \mathcal{L}(Q + \sigma^2 I)$ is convex in $Q \in \mathbb{S}^p_+$ by Theorem~\ref{thm:convex} (composition of a convex function with an affine mapping), and smooth on any compact subset.

\emph{Application.} By \citet{bhojanapalli2016dropping} (Theorem~2), when $d \geq r^*$, every second-order critical point $L^*$ of $g(L) = h(LL^\top)$ satisfies $g(L^*) = \min_{Q \in \mathbb{S}^p_+} h(Q)$.  Since every local minimum is a second-order critical point, every local minimum of $g$ is a global minimum.
\end{proof}

\subsection{Proof of Theorem~\ref{thm:consistency}}
\begin{proof}
The proof proceeds in three steps: uniform convergence of the empirical loss, identification of the unique population minimizer, and application of M-estimation theory.

\emph{Uniform convergence.} For any fixed $\Sigma^{-1} \in \mathbb{S}_+^p$, the empirical loss $\mathcal{L}_n(\Sigma^{-1}) = -\frac{1}{n}\sum_{i=1}^n \sum_k Y_{ik} \log P_{ik}(\Sigma^{-1})$ is an average of i.i.d.\ random variables (since each $(x_i, y_i)$ is drawn independently). We restrict attention to a compact set $\mathcal{C} = \{\Sigma^{-1} \in \mathbb{S}^p_+ : \|\Sigma^{-1}\|_F \leq C\}$ for some $C > 0$ chosen large enough that $\Sigma_0^{-1} \in \text{int}(\mathcal{C})$.

On $\mathcal{C}$, the per-sample loss function $\ell(x, y; \Sigma^{-1})$ is Lipschitz in $\Sigma^{-1}$ (because the log-softmax is Lipschitz in its arguments, and the logits are linear in $\Sigma^{-1}$ with coefficients bounded by the data and the means). The class of functions $\{\ell(\cdot; \Sigma^{-1}) : \Sigma^{-1} \in \mathcal{C}\}$ is therefore a Lipschitz-parametrized family over a bounded set, which forms a Glivenko-Cantelli class. By the uniform law of large numbers,
\[
\sup_{\Sigma^{-1} \in \mathcal{C}} \left| \mathcal{L}_n(\Sigma^{-1}) - \mathcal{L}_\infty(\Sigma^{-1}) \right| \xrightarrow{a.s.} 0,
\]
where the convergence also uses the consistency of $\hat{\mu}_k \to \mu_k^0$ and $\hat{\pi}_k \to \pi_k^0$.

\emph{Unique population minimizer.} Under the Gaussian model, the posterior class probability is $P(Y = k \mid X = x) = \text{softmax}_k(\delta(x; \Sigma_0^{-1}))$, where $\delta(x; \Sigma_0^{-1})$ denotes the discriminant evaluated at the true precision matrix. By the Gibbs inequality, the population cross-entropy
$\mathcal{L}_\infty(\Sigma^{-1}) = E\left[-\sum_k \mathbf{1}[Y=k] \log P_k(X; \Sigma^{-1})\right]$
is minimized when $P_k(x; \Sigma^{-1}) = P(Y=k \mid X=x)$ for almost all $x$.

This holds at $\Sigma^{-1} = \Sigma_0^{-1}$. Uniqueness follows because if $A \neq \Sigma_0^{-1}$ also achieved this, then $\delta_k(x; A) - \delta_j(x; A) = \delta_k(x; \Sigma_0^{-1}) - \delta_j(x; \Sigma_0^{-1})$ for almost all $x$ and all pairs $(k,j)$. Expanding this using the linearity of $\delta$ in the precision matrix, this requires $x^\top (A - \Sigma_0^{-1})(\mu_k^0 - \mu_j^0) = \frac{1}{2}(\mu_k^0 + \mu_j^0)^\top (A - \Sigma_0^{-1})(\mu_k^0 - \mu_j^0)$ for almost all $x$. Since $X$ has a density, this forces $(A - \Sigma_0^{-1})(\mu_k^0 - \mu_j^0) = 0$ for all pairs $(k,j)$. When the mean vectors $\{\mu_k^0\}$ span $\mathbb{R}^p$ (or more generally, when $K \geq 2$ and $\mu_1^0 \neq \mu_2^0$, combined with the full-rank covariance), this gives $A = \Sigma_0^{-1}$.

\emph{Consistency.} By standard M-estimation theory (see, e.g., \citet{van2000asymptotic}, Theorem 5.7): if the empirical objective converges uniformly to a limit that has a unique minimizer on a compact set, then the minimizers converge. That is, $\hat{\Sigma}_n^{-1} := \arg\min_{\Sigma^{-1} \in \mathcal{C}} \mathcal{L}_n(\Sigma^{-1}) \xrightarrow{P} \Sigma_0^{-1}$.

When $K - 1 \geq p$, the mean difference vectors $\{\mu_k^0 - \mu_1^0\}_{k=2}^K$ can span $\mathbb{R}^p$, in which case Step~2 shows $\Sigma_0^{-1}$ is the unique minimizer and the precision matrix estimator converges: $\hat{L}_n \hat{L}_n^\top + \sigma^2 I \xrightarrow{P} \Sigma_0^{-1}$. When $K - 1 < p$ (the typical regime), the minimizer is unique only on the discriminative subspace $\text{span}(\mu_k^0 - \mu_j^0 : k,j)$, but all minimizers yield the same class posteriors.

For the classification error, the predicted label is $\hat{y}(x) = \arg\max_k \delta_k(x; \hat{\Sigma}_n^{-1})$. Since the posteriors converge to the true posteriors (they depend only on the identified directions), $\hat{y}(x) \to y^*(x)$ (the Bayes classifier) for all $x$ not on the decision boundary (which has probability zero under continuous distributions). By the dominated convergence theorem, the misclassification rate converges to the Bayes error rate.
\end{proof}

\subsection{Proof of Theorem~\ref{thm:risk}}
\begin{proof}
We derive the excess risk bound in three steps: establishing a parametric rate for the precision matrix estimate, then converting it to a risk bound.

\emph{Local strong convexity.} With known class means $\mu_k^0$, the population cross-entropy $\mathcal{L}_\infty(\Sigma^{-1})$ depends on $\Sigma^{-1}$ through the logits $\delta_k(x) = x^\top \Sigma^{-1} \mu_k^0 - \frac{1}{2}(\mu_k^0)^\top \Sigma^{-1} \mu_k^0 + \log \pi_k^0$, which are affine in $\Sigma^{-1}$. The Hessian of $\mathcal{L}_\infty$ with respect to $\Sigma^{-1}$ evaluated at $\Sigma_0^{-1}$ is:
\[
\nabla^2 \mathcal{L}_\infty(\Sigma_0^{-1}) = E\left[\sum_{k=1}^K P_k^0(X)(1 - P_k^0(X)) \cdot a_k(X) a_k(X)^\top \right],
\]
where $P_k^0(x) = P(Y=k \mid X=x)$ is the true posterior and $a_k(x) = \text{vec}(x \mu_k^\top - \frac{1}{2}\mu_k \mu_k^\top)$ is the vectorized derivative of $\delta_k$ with respect to the entries of $\Sigma^{-1}$. When the classes are non-degenerate (i.e., $P_k^0(x) \in (0,1)$ on a set of positive measure and the vectors $a_k(X)$ span sufficient directions), this Hessian is positive definite. Hence $\mathcal{L}_\infty$ is locally strongly convex near $\Sigma_0^{-1}$.

\emph{Parametric estimation rate.} The factored model $\Sigma^{-1} = LL^\top$ with $L \in \mathbb{R}^{p \times d}$ has $pd$ free parameters. By the classical asymptotic theory of M-estimators, when the population objective is locally strongly convex and the empirical process satisfies a central limit theorem, the minimizer satisfies:
\[
\|\hat{L}_n \hat{L}_n^\top + \sigma^2 I - \Sigma_0^{-1}\|_F = \mathcal{O}_P\!\left(\sqrt{\frac{pd}{n}}\right).
\]
This is the standard parametric rate: the estimation error scales as the square root of the number of parameters divided by the sample size.

\emph{From estimation error to excess risk.} The misclassification risk $R(\Sigma^{-1}) = P(\hat{y}(X; \Sigma^{-1}) \neq Y)$ is minimized at $\Sigma_0^{-1}$, where it equals the Bayes risk $R^*$. Under Gaussian class-conditional distributions, the decision boundaries are hyperplanes, and the risk $R(\Sigma^{-1})$ can be expressed as an integral of the class-conditional densities over the complement of the predicted class region. Since the Gaussian density is smooth and the decision boundary moves smoothly with $\Sigma^{-1}$, the risk $R(\Sigma^{-1})$ is twice differentiable at $\Sigma_0^{-1}$.

Because $\Sigma_0^{-1}$ minimizes $R$ (it achieves the Bayes rate), the gradient $\nabla R(\Sigma_0^{-1}) = 0$. A second-order Taylor expansion gives:
\[
R(\hat{\Sigma}_n^{-1}) - R^* \leq C \cdot \|\hat{\Sigma}_n^{-1} - \Sigma_0^{-1}\|_F^2 = \mathcal{O}_P\!\left(\frac{pd}{n}\right),
\]
where $\hat{\Sigma}_n^{-1} = \hat{L}_n \hat{L}_n^\top + \sigma^2 I$ and $C$ depends on the curvature of $R$ at $\Sigma_0^{-1}$. This shows that LDA-GO with low-rank factorization achieves excess risk $\mathcal{O}_P(pd/n)$, which is tighter than the $\mathcal{O}_P(p^2/n)$ rate of the full-rank model with $d = p$.
\end{proof}

\end{document}